\documentclass[nofootinbib,superscriptaddress]{revtex4}
\usepackage{epsfig}
\usepackage{amsmath}
\usepackage{amsfonts}
\usepackage{amssymb}
\usepackage{color}
\usepackage{graphicx}
\usepackage{graphics}
\usepackage{hyperref}
\usepackage{epstopdf}
\usepackage{adjustbox,lipsum}

\begin{document}
\title{Triple Higgs Coupling as a Probe of the Twin-Peak Scenario}
\author{Amine Ahriche}
\email{aahriche@ictp.it}
\affiliation{Department of Physics, University of Jijel, PB 98 Ouled Aissa, DZ-18000 Jijel, Algeria}
\affiliation{The Abdus Salam International Centre for Theoretical Physics, Strada Costiera
11, I-34014, Trieste, Italy}
\affiliation{Fakult\"at f\"ur Physik, Universit\"at Bielefeld, 33501 Bielefeld, Germany}
\author{Abdesslam Arhrib}
\email{aarhrib@ictp.it}
\affiliation{Universit\'e AbdelMalek Essaadi, Facult\'{e} des Sciences et Techniques, B.P
416, Tangier, Morocco}
\author{Salah Nasri}
\email{snasri@uaeu.ac.ae}
\affiliation{Physics Department, UAE University, POB 17551, Al Ain, United Arab Emirates}

\begin{abstract}
In this letter, we investigate the case of a twin peak around the observed 125
GeV scalar resonance, using di-Higgs production processes at both LHC and
$e^{+}e^{-}$ Linear Colliders. We have shown that both at LHC and Linear
Collider the triple Higgs couplings play an important role to identify this
scenario; and also that this scenario can be distinguishable from any Standard
Model extension by extra massive particles which might modify the triple Higgs
coupling. We also introduce a criterion that can be used to ruled out the twin
peak scenario.

\end{abstract}
\maketitle

On July 2012, ATLAS and CMS collaborations \cite{ATLAS, CMS} have shown the
existence of a Higgs-like resonance around 125 \textrm{GeV} confirming the
cornerstone of the Higgs mechanism that predicted such particle long times
ago. All Higgs couplings measured so far seem to be consistent, to some
extent, with the Standard Model (SM) predictions. Moreover, in order to
establish the Higgs mechanism as responsible for the phenomena of electroweak
symmetry breaking one still needs to measure the self couplings of the Higgs
and therefore to reconstruct its scalar potential.

Recent measurements at the LHC show that there is still uncertainty on the
Higgs mass; $m_{h}=125.3\pm0.4(\text{stat}.)\pm0.5(\text{syst}.)$ \textrm{GeV}
for CMS \cite{CMS2G} and $m_{h}=125.0\pm0.5$ \textrm{GeV} for ATLAS
\cite{ATLAS2G} from the diphoton channel and $m_{h}=125.5\pm0.37(\text{stat}%
.)\pm0.18(\text{syst}.)$ \textrm{GeV} from combined channels. Despite this
relatively large uncertainty, a scenario of two degenerate scalars around
125.5 \textrm{GeV} resonance is neither excluded nor confirmed \cite{TP}.

In the twin peak scenario (\textrm{TPS}); it is assumed that there are two
scalars $h_{1,2}$ with almost degenerate masses around 125 \textrm{GeV}. To
our knowledge, there is no indication from experimental data which disfavor
this scenario. The couplings of the twin peak Higgs to SM particles
$g_{h_{i}XX}$ are simply scaled with respect to SM rate by $\cos\theta$ (for
$h_{1}$) and $\sin\theta$ (for $h_{2}$), where $\theta$ is a mixing angle,
such that we have the following approximate sum rule:%
\begin{equation}
g_{h_{1}f\bar{f}}^{2}+g_{h_{2}f\bar{f}}^{2}\simeq g_{h_{SM}f\bar{f}}%
^{2},~g_{h_{1}VV}^{2}+g_{h_{2}VV}^{2}\simeq g_{h_{SM}VV}^{2}, \label{sr}%
\end{equation}
where $f$ can be any of the SM fermions and $V=W,Z$ vector boson. In
fact, the branching ratios of the Higgs to SM particles are SM-like
only if the Higgs invisible is very suppressed or kinematically
forbidden as will be considered in our example. Consequently, the
single Higgs production such as gluon-gluon fusion at LHC,
Higgs-strahlung, Vector Boson Fusions, and $t\bar{t}H$ at LHC and
$e^{+}e^{-}$ Linear Colliders (LC) will obey the same sum rule. The
summation of event numbers (both for production and decay) of the
two possible cases will be identical to SM case since
$\cos^{2}\theta +\sin^{2}\theta=1$. However, for processes with
di-Higgs final states ($pp(e^{-}e^{+})\rightarrow hh+X$), the triple
Higgs couplings may play an important role, and therefore these
processes can be useful to distinguish between the cases of one
scalar or two degenerate ones around the observed 125 \textrm{GeV}
resonance.

It is well known that the triple Higgs couplings can be, in principle,
measured directly at the LHC with high luminosity option through double Higgs
production $pp\rightarrow gg\rightarrow hh$ \cite{double1}. Such measurement
is rather challenging at the LHC, and for this purpose several parton level
analysis have been devoted to this process. It turns out that $hh\rightarrow
b\bar{b}\gamma\gamma$ \cite{double3}, $hh\rightarrow b\bar{b}\tau^{+}\tau^{-}$
\cite{double3,double0} and $hh\rightarrow b\bar{b}W^{+}W^{-}$
\cite{double0,ww} final states are very promising for High luminosity.
Recently, CMS report a preliminary result on the search for resonant di-Higgs
production in $b\bar{b}\gamma\gamma$ channel \cite{CMS-hh}. \newline The LC
has also the capability of measuring with better precision: the Higgs mass and
some of the Higgs couplings together with the self coupling of the Higgs
\cite{ILC}. Using recoil technique for the Higgs-strahlung process, the Higgs
mass can be measured with an accuracy of about 40 \textrm{MeV} \cite{ILC}. We
note that at LHC with high luminosity we can measure the Higgs mass with about
100 \textrm{MeV} uncertainty which is quite comparable to $e^{+}e^{-}$
colliders. The triple Higgs coupling can be extracted from $e^{+}%
e^{-}\rightarrow Zh^{\ast}\rightarrow Zhh$ at 500 \textrm{GeV} and even better
from $e^{+}e^{-}\rightarrow\nu\overline{\nu}h^{\ast}\rightarrow\nu
\overline{\nu}hh$ at $\sqrt{s}>800$ \textrm{GeV}. In this regards, the LHC and
$e^{+}e^{-}$ LC measurements are complementary \cite{Weiglein:2004hn}.

In Ref. \cite{Gunion}, the authors have provided a tool to distinguish the
two-degenerate states scenario from the single Higgs one. The approach of
\cite{Gunion} applies only to models which enjoy modifications of
$h\rightarrow\gamma\gamma$ rate with respect to the SM. However, according to
the latest experimental results, both for ATLAS and CMS the di-photon channel
seem to be rather consistent with the SM \cite{CMS2G,ATLAS2G}. In this work we
propose a new approach to distinguish the TPS. This approach is based on the
di-Higgs production which is sensitive to the triple Higgs coupling, that is
modified in the majority of SM extensions. \newline Here, as an example, we
consider, the Two-Singlets Model proposed in \cite{TSM}, where the SM is
extended with two real scalar fields $S_{0}$ and $\chi_{1}$; each one is odd
under a discrete symmetry $\mathbb{Z}_{2}^{(0)}$ and $\mathbb{Z}_{2}^{(1)}$
respectively. The field $\chi_{1}$ has a non vanishing vacuum expectation
value, which breaks $\mathbb{Z}_{2}^{(1)}$ spontaneously, whereas,
$\left\langle S_{0}\right\rangle =0$; and hence, $S_{0}$ is a dark matter
candidate. Both fields are SM gauge singlets and hence can interact with the
'visible'\ particles only via the Higgs doublet $H$. The spontaneous breaking
of the electroweak and the $\mathbb{Z}_{2}^{(1)} $ symmetries introduces the
two vacuum expectation values $\upsilon$ and $\upsilon_{1}$ respectively. The
physical Higgs $h_{1}$ and\ $h_{2}$, with masses $m_{1}$ and $m_{2}\gtrsim
m_{1}$, are related to the excitations of the neutral component of the SM
Higgs doublet field, Re$(H^{(0)})$, and the field $\chi_{1}$ through rotation
with a mixing angle $\theta$ and, with a specific choice in the parameter
space, could give rise to two degenerate scalars around 125 \textrm{GeV}. In
what follows, we denote by $c=\cos\theta$ and $s=\sin\theta$. The quartic and
triple couplings of the physical fields $h_{i}$ are given in the appendices in
\cite{Arh}. \newline In our analysis we require that \footnote{Actually, we
considered that all quartic couplings to be of order unity; and the singlet
vev $\upsilon_{1}=\left\langle \chi_{1}\right\rangle =20\sim2000$
$\mathrm{GeV.}$}: \textit{(i)} all the dimensionless quartic couplings to be
$\ll4\pi$ for the theory to remain perturbative, \textit{(ii)} the two scalar
eigenmasses should be in agreement with recent measurements
\cite{CMS2G,ATLAS2G}: we have checked that for the Two-Singlets model, the
splitting between $m_{1}$ and $m_{2}$ could be of the order of 40
\textrm{MeV}. \textit{(iii)} the ground state stability to be ensured; and
\textit{(iv)} we allow the DM mass $m_{0}$ to be as large as 1 \textrm{TeV}.
\newline In our work, we consider di-Higgs production processes at the LHC and
$e^{+}e^{-}$ LC, whose values of the cross section could be significant,
namely, $\sigma^{LHC}(hh)$ and $\sigma^{LHC}(hh+t\bar{t})$ at 14 \textrm{TeV};
$\sigma^{LC}(hh+Z)$ at 500 \textrm{GeV}\ and $\sigma^{LC}(hh+E_{miss})$ at 1
\textrm{TeV}. All these processes include, at least, one Feynman diagram with
triple Higgs coupling. For the \textrm{TPS}, the total cross section gets
contributions from the final states $h_{1}h_{1}$, $h_{1}h_{2}$ and $h_{2}%
h_{2}$. Therefore the quantity to be compared with the standard scenario can
be expressed as:
\begin{equation}
\sigma^{\mathrm{TPS}}\left( hh+X\right) =\sigma\left(
h_{1}h_{1}+X\right) +2\sigma\left( h_{1}h_{2}+X\right) +\sigma\left(
h_{2}h_{2}+X\right) ,
\label{TPS}%
\end{equation}
which can be parameterized as:
\begin{equation}
\sigma^{\mathrm{TPS}}=\sigma_{aa}r_{1}+\sigma_{ab}r_{2}+\sigma_{bb},
\end{equation}
with $\sigma_{aa}+\sigma_{ab}+\sigma_{bb}=\sigma^{SM}\left(
hh+X\right) $ and $\sigma_{aa}$, $\sigma_{bb}$ and $\sigma_{ab}$
correspond to the cross section contributions coming from triple
Higgs diagrams ($a$), non-triple Higgs diagrams ($b$) and the
interference term in the amplitude, respectively. The coefficients
$r_{i}$ are dimensionless parameters, that receive contributions
from the final states $h_{i}h_{j}$, which depend on the mixing angle
$\theta$ and the Higgs triple couplings $\lambda_{ijk}^{(3)} $.

\begin{figure}[t]
\begin{centering}
\includegraphics[width=7cm,height=5.5cm]{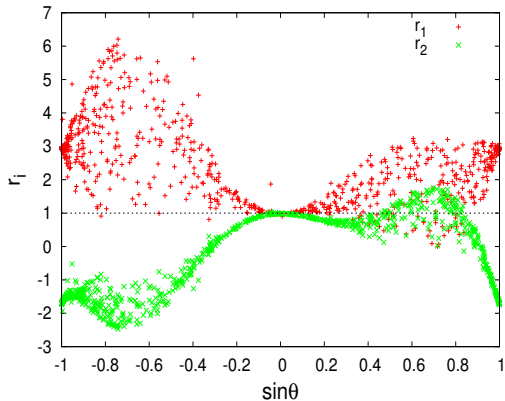}
\par\end{centering}
\caption{\textit{Numerical values of the parameters $r_{i}$ in (\ref{ri}) for
600 benchmarks that fulfill the above mentioned requirements.}}%
\label{rri}%
\end{figure}

In the \textrm{TPS}, the amplitudes for di-Higgs production processes have SM
Feynman diagrams where the the Higgs field h is replaced by $h_{i}$. To
compute the parameters $r_{i}$, we first estimate how does each amplitude get
modified with respect to the corresponding SM one for each case $h_{i}h_{j}$.
For example, in the case of $h_{1}h_{1}$ production, there are two types of
diagrams: (1) The ones that involve triple scalar interactions $h_{1}%
h_{1}h_{1}$ and $h_{2}h_{1}h_{1}$, with couplings equal to the one of a SM
times a factor of $c\lambda_{111}^{(3)}/\lambda_{hhh}^{SM}$ and $s\lambda
_{112}^{(3)}/\lambda_{hhh}^{SM}$, respectively. We denote the total amplitude
of these two contributions by $\mathcal{M}_{(a)}$. (2) The ones with no triple
Higgs couplings. Their amplitude, denoted by $\mathcal{M}_{(b)}$, is given by
the one of the SM scaled by a factor of $c^{2}$. Therefore, the amplitudes
$\mathcal{M}_{(a,b)}$ (where $a$ ($b$) stand for triple Higgs (non-triple
Higgs) Feynman diagrams) for the di-Higgs production can be written in terms
of their corresponding SM values as:
\[%
\begin{tabular}
[c]{lll}\hline\hline $h_{1}h_{1}:$ & &
$\mathcal{M}_{(a)}=[(c\lambda_{111}^{(3)}+s\lambda
_{112}^{(3)})/\lambda_{hhh}^{SM}]\mathcal{M}_{(a)}^{SM},$\\
& & $\mathcal{M}_{(b)}=c^{2}\mathcal{M}_{(b)}^{SM},$\\\hline
$h_{2}h_{2}:$ & & $\mathcal{M}_{(a)}=[(c\lambda_{122}^{(3)}+s\lambda
_{222}^{(3)})/\lambda_{hhh}^{SM}]\mathcal{M}_{(a)}^{SM},$\\
& & $\mathcal{M}_{(b)}=s^{2}\mathcal{M}_{(b)}^{SM},$\\\hline
$h_{1}h_{2}:$ & & $\mathcal{M}_{(a)}=[(c\lambda_{112}^{(3)}+s\lambda
_{122}^{(3)})/\lambda_{hhh}^{SM}]\mathcal{M}_{(a)}^{SM},$\\
& & $\mathcal{M}_{(b)}=cs\mathcal{M}_{(b)}^{SM},$\\\hline\hline
\end{tabular}
\ \
\]
where $\lambda_{hhh}^{SM}$ is the SM triple Higgs coupling calculated at
one-loop. Then the parameters $r_{i}$ are given by:
\begin{gather}
r_{1}=\left\{
c^{2}[\lambda_{111}^{(3)2}+\lambda_{122}^{(3)2}+2\lambda
_{112}^{(3)2}]+s^{2}[\lambda_{112}^{(3)2}+\lambda_{222}^{(3)2}+2\lambda
_{122}^{(3)2}]+2cs[\lambda_{111}^{(3)}\lambda_{112}^{(3)}+2\lambda_{112}%
^{(3)}\lambda_{122}^{(3)}+\lambda_{122}^{(3)}\lambda_{222}^{(3)}]\right\}
/\left( \lambda_{hhh}^{SM}\right)^{2},\nonumber\\
r_{2}=\{c^{3}\lambda_{111}^{(3)}+3c^{2}s\lambda_{112}^{(3)}+3cs^{2}%
\lambda_{122}^{(3)}+s^{3}\lambda_{222}^{(3)}\}/\lambda_{hhh}^{SM}. \label{ri}%
\end{gather}
Thus, the values of $r_{i}$ quantify by how much each di-Higgs process
deviates from the SM case. In Fig. \ref{rri}, we show the parameters $r_{i}$
as a function of $\sin\theta$ for about 600 chosen sets of the model
parameters within the condition (\ref{sr}). We see that for very small mixing angle $r_{i}$'s are
approximately equal to unity, while for $\sin\theta>0.8$ and $\sin\theta
<-0.2$, the parameter $r_{1} $ becomes larger than unity and $r_{2}$ acquires
negative values. This behavior could lead to an enhancement/reduction to the
cross section depending on the sign of the interference contribution,
$\sigma_{ab}$, to the total cross section. This means that the measurement of
the following ratio:
\begin{equation}
\xi\left( hh+X\right) =\frac{\sigma^{\mathrm{TPS}}\left( pp(e^{-}%
e^{+})\rightarrow hh+X\right) }{\sigma^{SM}\left(
pp(e^{-}e^{+})\rightarrow
hh+X\right) }, \label{xi}%
\end{equation}
could be very useful to confirm or exclude this scenario based on
the deviation of any of the parameters $r_{i}$ from unity. For
instance, the ratio $\xi\left( hh+X\right) $ can deviate from unity
if the SM is extended with massive particles (SM+MP) that couple to
the Higgs doublet and contribute to the triple Higgs coupling as
well the Higgs mass. In this case, $r_{1}=\left( 1+\Delta\right)
^{2}$ and $r_{2}=1+\Delta$, where $\Delta$ represents the relative
enhancement of the triple Higgs coupling due to SM+MP. As we will
show later, our considered scenario for small or large mixing could
be distinguished from the case of SM+MP by combining the ratio
(\ref{xi}) for different processes.

In Table \ref{T2}, we give the values of $\sigma_{aa}$, $\sigma_{ab}$\ and
$\sigma_{bb}$\ for the corresponding di-Higgs production processes. We note
that their contributions to the LHC process $pp\rightarrow hh$ and to the LC
one $e^{+}e^{-}\rightarrow Zhh$ seem to be uncorrelated, which makes the Higgs
triple coupling useful to probe this scenario and distinguish it from (SM+MP).
\begin{table}[h]
\begin{adjustbox}{max width=0.8\textwidth}
\begin{tabular}{c|c|c|c|c|}
\cline{2-5} & $\sigma_{aa}~(fb)$ & $\sigma_{ab}~(fb)$ & $\sigma
_{bb}~(fb)$ & $\sigma^{SM}~(fb)$ \\ \hline
\multicolumn{1}{|c|}{$hh$} & $9.66$ & $-49.9$ & $70.1$ & $29.86$ \\
\hline
\multicolumn{1}{|c|}{$hh+t\bar{t}$} & $3.3164\times 10^{-2}$ & $0.13952$ & $%
0.84731$ & $1.02$ \\ \hline \multicolumn{1}{|c|}{$hh+Z$} &
$9.0206\times 10^{-3}$ & $4.6999\times 10^{-2} $ & $9.005\times
10^{-2}$ & $0.14607$ \\ \hline
\multicolumn{1}{|c|}{$hh+E_{miss}$} & $5.1631\times 10^{-2}$ & $-0.20867$ & $%
0.29708$ & $0.14004$ \\ \hline
\end{tabular}
\end{adjustbox}
\caption{\textit{Different contributions to the considered processes cross
sections. Numbers for LHC are taken from \cite{Spira:1995mt} at NLO.} }%
\label{T2}%
\end{table}

\begin{figure}[t]
\begin{centering}
\includegraphics[width=7cm,height=5.5cm]{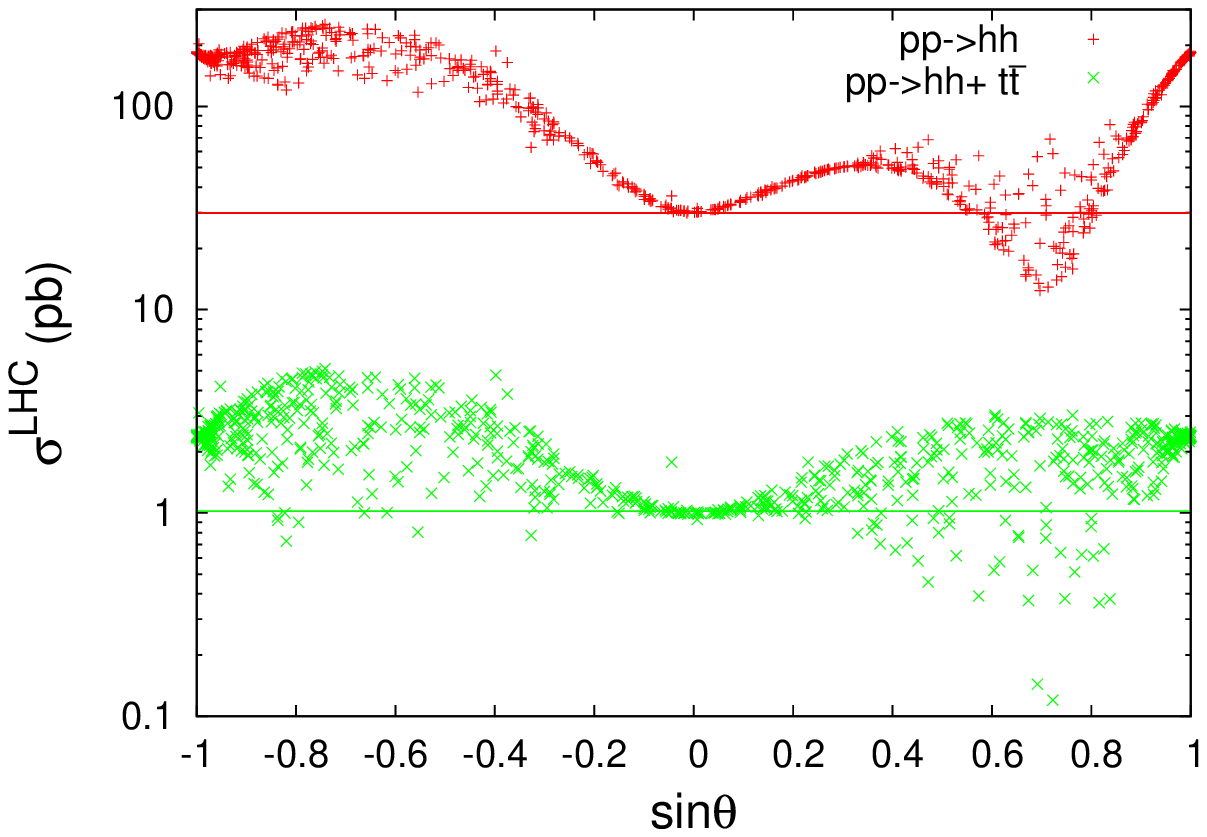}~\includegraphics[width=7cm,height=5.5cm]{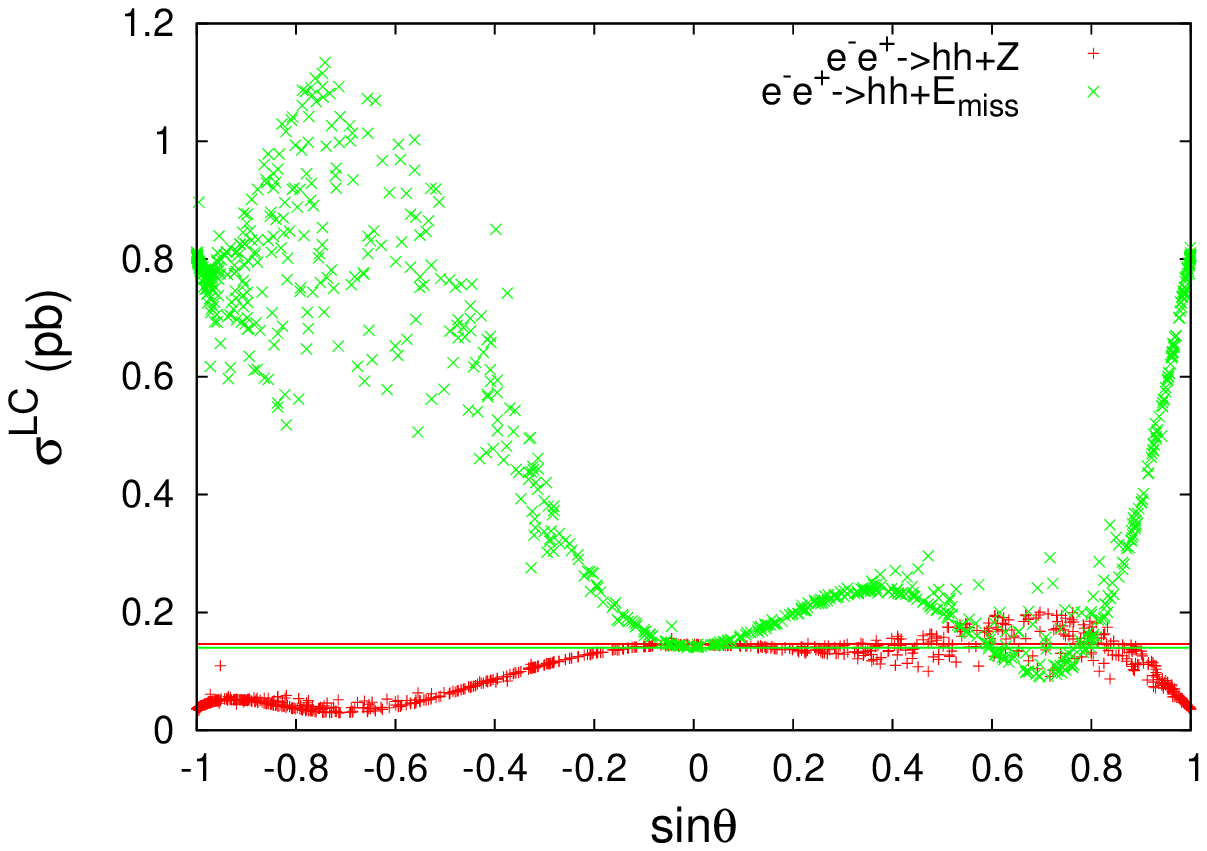}
\par\end{centering}
\caption{\textit{The cross section values (\ref{TPS}) for the di-Higgs
production processes for the 600 benchmarks used previously. The solid lines
correspond to the SM cross sections.}}%
\label{cs}%
\end{figure}

\begin{figure}[h]
\begin{centering}
\includegraphics[width=7.8cm,height=5.5cm]{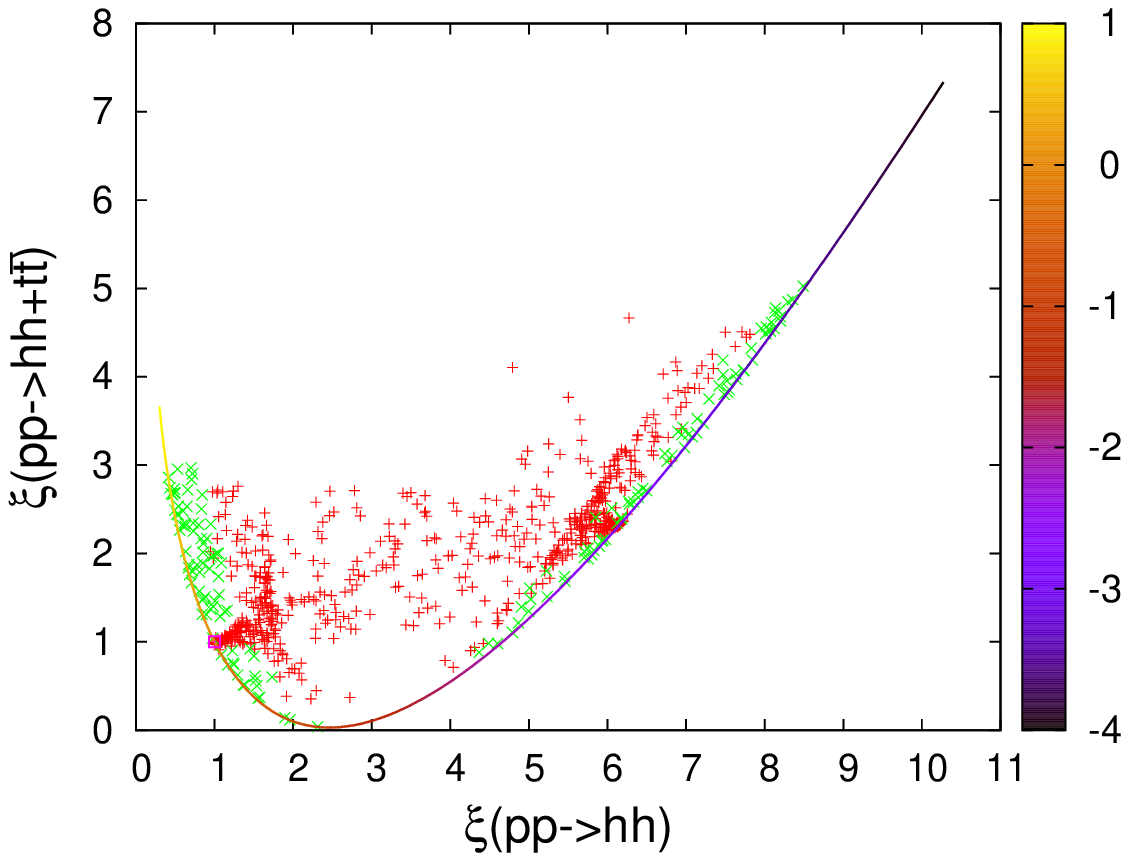}~\includegraphics[width=7.8cm,height=5.5cm]{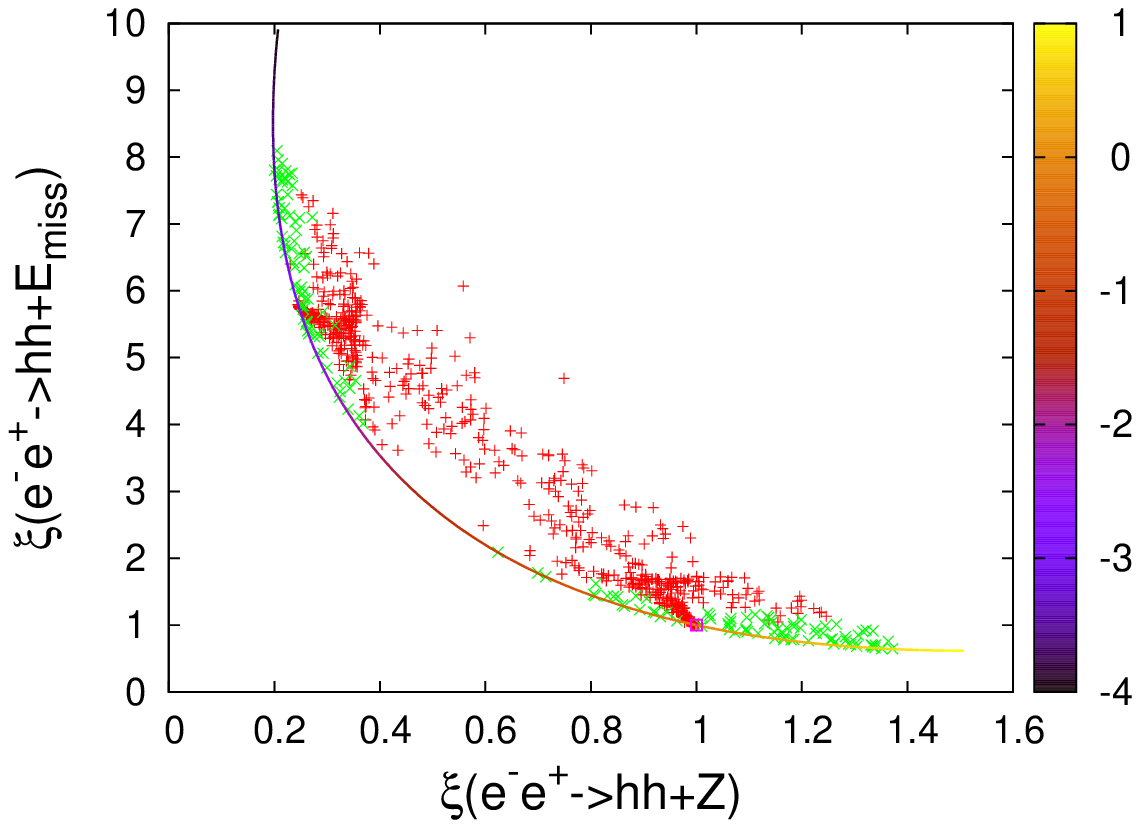}
\par\end{centering}
\caption{\textit{The ratios $\xi$ given in (\ref{xi}) for the di-Higgs
production processes for the 600 benchmark used previously. The green
benchmarks correspond to the large mixing case where $0.35<cos^{2}\theta
<0.65$, and the blue point represents the SM; and the solid curve represents
the case of a SM extension, where the new physics affects the triple Higgs
coupling as $\lambda_{hhh}=\lambda_{hhh}^{SM}(1+\Delta)$; and the value of the
relative enhancement $\Delta$ can be read from the palette.}}%
\label{rat}%
\end{figure}

For the benchmarks considered previously in Fig. \ref{rri}, we illustrate in
Fig. \ref{cs} the production cross section of di-Higgs at $e^{+}e^{-}$ LC and
LHC and in Fig. \ref{rat} the ratio $\xi$. As it can be seen, in the TPS, the
cross section of the processes $pp\rightarrow hh$, $pp\rightarrow hh+t\bar{t}$
and $e^{-}e^{+}\rightarrow hh+E_{miss}$ are mostly enhanced, while for
$e^{-}e^{+}\rightarrow hh+Z$ it is enhanced just for the mixing values
$0.5<\sin\theta<0.8$.

Now let us discuss the possibility of disentangling the \textrm{TPS} from the
SM+MP. It is clear from Fig. \ref{rat} that for both LHC and LC processes with
large mixing, $0.35<\cos^{2}\theta<0.65$, the TPS may coincide with SM+MP.
However, for non-maximal mixing values the TPS is clearly different than SM+MP
where all benchmarks have the following feature
\begin{equation}
\xi_{1}^{TPS}+\xi_{2}^{TPS}>\xi_{1}^{SM+MP}\left( \Delta\right)
+\xi
_{2}^{SM+MP}\left( \Delta\right) , \label{crit}%
\end{equation}
where $\xi_{i}^{TPS}$ the ratio in (\ref{xi}) for any LHC or LC
processes and $\xi_{i}^{SM+MP}\left( \Delta\right) $ is the same
ratio due the existence of massive particles. Therefore, when
measuring the quantities (\ref{xi}) for both the LHC and
$e^{+}e^{-}$ LC processes, and one finds that the criterion
(\ref{crit}) is not fulfilled, then it is a certain exclusion
exclusion for this scenario. In case where the criterion
(\ref{crit}) is fulfilled, detailed analysis is required for in
order to identify the mixing angle, the parameters
$\mathbf{r}_{\mathbf{i}}$ and therefore the Higgs triple couplings.
In fact, by studying all the di-Higgs production channels at both
LHC and $e^{+}e^{-}$ LC one not only confirm/exclude this scenario,
but also distinguished it from models where only one type of
processes gets modified by new physics such as: it manifests as new
sources of missing energy in $e^{-}e^{+}\rightarrow hh+E_{miss}$
\cite{Ahriche:2014xra}, new colored scalar singlets contribution to
$pp\rightarrow hh$ (or $hh+t\bar{t}$) \cite{Kribs:2012kz}, or the
presence of a heavy resonant Higgs \cite{resonant}.

In order to show whether this scenario can be tested at colliders, we consider
three benchmarks that may be distinguished from SM+MP (i.e., three red points
from Fig. \ref{rat}), and compare the di-Higgs distribution (of the di-Higgs
invariant mass as an example) with the SM one. The corresponding values of
ratios $r_{i}$ and $\xi_{i}$ are given in Table \ref{BB}, and in Table
\ref{en}, we present the expected number of events at both the LHC and LC. We
see that for benchmark $B_{2}$, the events number is significantly larger than
the SM for the channels $pp\rightarrow2b2\tau$ at the LHC and $e^{-}%
e^{+}\rightarrow4b+E_{miss}$ at LC's, while it is reduced for the processes
$pp\rightarrow4b+t\bar{t}$ and $e^{-}e^{+}\rightarrow4b+Z$. For benchmark
$B_{1}$, the events number of the processes $pp\rightarrow2b2\tau$ and
$e^{-}e^{+}\rightarrow4b+E_{miss}$ is SM-like but it is reduced for the
processes $pp\rightarrow4b+t\bar{t}$ and $e^{-}e^{+}\rightarrow4b+Z$. For
benchmark $B_{3}$, the events number is reduced for the considered processes.

\begin{table}[h]
\begin{adjustbox}{max width=0.8\textwidth}
\begin{tabular}{cccc}
\hline\hline & $B_{1}$ & $B_{2}$ & $B_{3}$ \\ \hline $\sin \theta $
& $0.53555$ & $0.90126$ & $-0.39802$ \\
\hline $r_{1}$ & $2.95386$ & $,2.88466$ & $5.62286$\\ \hline $r_{2}$
& $1.31634$ & $0.28189$ &
$-1.26011$ \\
\hline $\xi \left( hh\right) $ & $1.10345$ & $2.80975$ & $6.27248$
\\ \hline
$\xi \left( hh+t\bar{t}\right) $ & $2.69728$ & $2.51821$ & $4.66603$ \\
\hline $\xi \left( hh+Z\right) $ & $ 1.22243$ & $0.88532$ &
$0.55827$
\\ \hline
$\xi \left( hh+E_{miss}\right) $ & $1.24900$ & $2.76488$ & $6.07213$ \\
\hline\hline
\end{tabular}
\end{adjustbox}
\caption{\textit{Different values of the ratios (\ref{ri}) and (\ref{xi}) for
the three chosen benchmarks.}}%
\label{BB}%
\end{table}

\begin{table}[h]
\begin{adjustbox}{max width=0.8\textwidth}
\begin{tabular}
[c]{cccccc}\hline\hline $Events~number$ & $channel$ & $SM$ & $B_{1}$
& $B_{2}$ & $B_{3}$\\\hline
$pp\rightarrow hh$ & $4b$ & $966.75$ & $1066.8$ & $2716.3$ & $6063.9$\\
& $2b2\tau$ & $106.70$ & $117.74$ & $299.8$ & $669.27$\\
& $2b2\gamma$ & $3.89$ & $4.29$ & $10.93$ & $24.4$\\\hline
$pp\rightarrow hh+t\bar{t}$ & $4b$ & $33.02$ & $89.06$ & $83.15$ &
$154.07$\\\hline $e^{-}e^{+}\rightarrow hh+Z$ & $4b$ & $23.65$ &
$28.91$ & $20.94$ & $13.2$\\\hline $e^{-}e^{+}\rightarrow
hh+E_{miss}$ & $4b$ & $45.34$ & $56.63$ & $125.36$ &
$275.31$\\\hline\hline
\end{tabular}
\end{adjustbox}
\caption{\textit{The events number for the different processes within the
luminosity values mentioned above for the SM and the benchmarks shown in Table
\ref{BB}.}}%
\label{en}%
\end{table}

\begin{figure}[h]
\begin{centering}
\includegraphics[width=8cm,height=5.5cm]{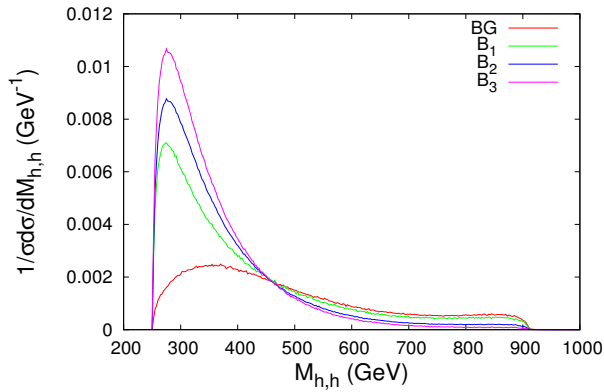}
\par\end{centering}
\caption{\textit{Normalized di-Higgs invariant mass distribution for the
process $e^{-}e^{+}\rightarrow hh+E_{miss}$ for the background (BG) and the
considered benchmarks in Table \ref{BB}.}}%
\label{Mhh}%
\end{figure}

In Fig. \ref{Mhh}, we illustrate the di-Higgs invariant mass distribution
($M_{h,h}$) for the process $e^{-}e^{+}\rightarrow hh+E_{miss}$. Clearly, the
\textrm{TPS} can be easily distinguished from the SM, especially in the case
of non-maximal mixing. However, the full confirmation of the \textrm{TPS}
requires the enlargement of the investigation by taking into account other
di-Higgs production channels such as $hhjj$, $hhW^{\pm}$, $hhZ$ and $hhtj$ at
the LHC \cite{Frederix:2014hta} and the $e^{+}e^{-}$ LC \cite{ILC}.

In conclusion, we have investigated the case of twin-peak at the 125 GeV
observed scalar resonance by considering different di-Higgs production
processes at both LHC and $e^{+}e^{-}$ LC. We have introduced a criterion
whose violation excludes the TPS scenario, otherwise this scenario can be
surely distinguished from the SM and SM extended by massive fields in case of
non-maximal mixing.

Last but not least, we should note that this scenario could be
realized within SM +(real/complex) singlet scalar, or any larger
scalar field content. This includes neutral or charged scalars that
are members any multiplets, where two degenerate scalar eigenstates
$h_{1,2}$ at 125 \textrm{GeV}, do couple to the SM gauge fields and
fermions by more than $\sim$90\%, i.e., the sum rule (\ref{sr}) is
fulfilled \footnote{In the 2HDM, twin pick scenario has been studied
in \cite{Ferreira:2012nv}, but the study concentrated only on the
diphoton channel. According to this study \cite{Ferreira:2012nv},
this scenario is not ruled out.}. If the measurement of di-Higgs
processes at LHC and/or $e^{+}e^{-}$ LC turn out to be consistent
with SM predictions, then it will be very challenging to distinguish
the \textrm{TPS} scenario.

If the measurement of the couplings $hf\bar{f}$ and $hVV$ become
much more precise from the future experiment data, it may be
possible that one could be sensitive to the radiative corrections
effect to these couplings. Such radiative corrections to $hf\bar{f}$
and $hVV$ couplings in a variety of extended Higgs sector have been
evaluated in \cite{Kanemura:2014dja,Aoki:2012jj,Arhrib:2003ph}.
These one-loop effects are of the order of 2-10\% and even more in
some special cases. The present LHC measurements are not yet
sensitive to such effects.

\subsection*{Acknowledgments}

We would like to thank A. Djouadi and R. Santos for the valuable comments; and
E.~Vryonidou for sharing with us her code and for many useful discussions. A.
Ahriche is supported by the Algerian Ministry of Higher Education and
Scientific Research under the CNEPRU Project No. D01720130042; and partially
by DAAD and ICTP. A. Arhrib is supported in part by the Moroccan Ministry of
Higher Education and Scientific Research: "projet des domaines prioritaires de
la recherche scientifique et du developpement technologique".

\end{document}